\documentclass[a4paper,11pt]{article}
\usepackage{pos}
\usepackage[braket,qm]{qcircuit}

\usepackage{bbm}

\newcommand{\cn}{\ensuremath{\mathbb{C}}}
\newcommand{\rn}{\ensuremath{\mathbb{R}}}
\newcommand{\zn}{\ensuremath{\mathbb{Z}}}

\newcommand{\Pp}{\ensuremath{\mathcal{P}}}
\newcommand{\Sp}{\ensuremath{\mathcal{S}}}
\newcommand{\Dp}{\ensuremath{\mathcal{D}}}
\newcommand{\Mp}{\ensuremath{\mathcal{M}}}
\newcommand{\Op}{\ensuremath{\mathcal{O}}}
\renewcommand{\d}{{\ensuremath{\partial}}}
\renewcommand{\theta}{\vartheta}
\renewcommand{\phi}{\varphi}
\newcommand{\eps}{\varepsilon}
\renewcommand{\l}{\left}
\renewcommand{\r}{\right}
\newcommand{\CNOT}{\ensuremath{\mathrm{CNOT}}}
\newcommand{\prob}{\ensuremath{\mathrm{prob}}}
\newcommand{\anc}{\ensuremath{\mathrm{ancilla}}}
\newcommand{\init}{\ensuremath{\mathrm{init}}}
\newcommand{\vol}{\ensuremath{\mathrm{vol}}}

\usepackage{xcolor}

\title{Dimensional Expressivity Analysis, best-approximation errors, and automated design of parametric quantum circuits}
 \ShortTitle{Dimensional Expressivity Analysis}

\author[a,b]{Lena~Funcke}
\author*[c,d]{Tobias~Hartung}
\author[e]{Karl~Jansen}
\author[d]{Stefan~K\"uhn}
\author[e,f]{Manuel~Schneider}
\author[e,g]{Paolo~Stornati}

\affiliation[a]{Center for Theoretical Physics, Co-Design Center for Quantum Advantage, and NSF AI Institute for Artificial Intelligence and Fundamental Interactions, Massachusetts Institute of Technology, 77 Massachusetts Avenue, Cambridge, MA 02139, USA}

\affiliation[b]{Perimeter Institute for Theoretical Physics, 31 Caroline Street North, Waterloo, ON N2L 2Y5, Canada}

\affiliation[c]{Department of Mathematical Sciences, University of Bath, 4 West, Claverton Down, Bath, BA2 7AY, UK}

\affiliation[d]{Computation-based  Science  and  Technology  Research  Center, The  Cyprus  Institute,  20  Kavafi  Street,  2121  Nicosia, Cyprus}

\affiliation[e]{Deutsches Elektronen-Synchrotron DESY, Platanenallee 6, 15738 Zeuthen}

\affiliation[f]{Institut für Physik, Humboldt-Universität zu Berlin, Newtonstraße 15, 12489 Berlin, Germany}

\affiliation[g]{ICFO, The Barcelona Institute of Science and Technology, Av. Carl Friedrich Gauss 3, 08860 Castelldefels (Barcelona), Spain}

\emailAdd{lfuncke@mit.edu}
\emailAdd{tobias.hartung@desy.de}
\emailAdd{s.kuehn@cyi.ac.cy}
\emailAdd{karl.jansen@desy.de}
\emailAdd{manuel.schneider@desy.de}
\emailAdd{paolo.stornati@desy.de}

\abstract{
The design of parametric quantum circuits (PQCs) for efficient use in variational quantum simulations (VQS) is subject to two competing factors. On one hand, the set of states that can be generated by the PQC has to be large enough to contain the solution state. Otherwise, one may at best find the best approximation of the solution restricted to the states generated by the chosen PQC. On the other hand, the PQC should contain as few parametric quantum gates as possible to minimize noise from the quantum device. Thus, when designing a PQC one needs to ensure that there are no redundant parameters. The dimensional expressivity analysis discussed in these proceedings is a means of addressing these counteracting effects. Its main objective is to identify independent and redundant parameters in the PQC. Using this information, superfluous parameters can be removed and the dimension of the space of states that are generated by the PQC can be computed. Knowing the dimension of the physical state space then allows us to deduce whether or not the PQC can reach all physical states. Furthermore, the dimensional expressivity analysis can be implemented efficiently using a hybrid quantum-classical algorithm. This implementation has relatively small overhead costs both for the classical and  quantum part of the algorithm and could therefore be used in the future for on-the-fly circuit construction. This would allow for optimized circuits to be used in every loop of a VQS rather than the same PQC for the entire VQS. These proceedings review and extend work in~\cite{Funcke2021a,Funcke2021b}. \\

Preprint number: MIT-CTP/5349 
}

\FullConference{%
 The 38th International Symposium on Lattice Field Theory, LATTICE2021
  26th-30th July, 2021
  Zoom/Gather@Massachusetts Institute of Technology
}

\begin{document}
\maketitle

\section{Introduction}\label{sec:intro}

The fast development of noisy intermediate-scale quantum (NISQ) computers~\cite{Preskill2018} builds the foundation for a large class of computational problems that cannot be solved efficiently with classical computers to be addressed with quantum devices. Such applications of quantum computing include machine learning~\cite{Biamonte2017}, finance~\cite{Orus2019}, various optimization problems~\cite{Montanaro2016,Brandao2017}, as well as physics. A great advantage of quantum computation in physics is that it can circumvent the sign problem which prevents Monte Carlo simulations of many strongly correlated quantum-many body problems~\cite{Troyer2005}. Although current quantum hardware is of limited size and suffers from a considerable level of noise, NISQ devices have already successfully demonstrated their ability to outperform classical computers~\cite{Arute2019,Zhong2020} and techniques for mitigating the effects of noise are rapidly developing~\cite{Temme2017,Kandala2017,Endo2018,YeterAydeniz2020,Funcke2020}.

An important class of algorithms designed for NISQ devices are variational quantum simulations~(VQSs)~\cite{Peruzzo2014,McClean2016}. These are hybrid quantum-classical algorithms for solving optimization problems and make use of parametric quantum circuits (PQCs), i.e., quantum circuits composed of parameter dependent gates. VQSs generally consist of a classical feedback loop optimizer that aims to minimize a given cost function which can be evaluated efficiently using the quantum device as a co-processor. In many cases, the cost function is related to the energy of the quantum state prepared on the quantum device by the PQC at a given set of parameters. As such, VQSs are naturally attuned to applications in quantum many-body systems in quantum chemistry~\cite{Peruzzo2014,Wang2015,Kandala2017,Hempel2018}, as well as in quantum mechanics and quantum field theory~\cite{Kokail2018,Hartung2018,Jansen2019,Paulson2020,Haase2020}.

Since VQSs intrinsically depend on the design of PQCs, finding a good or optimal PQC is paramount. This naturally leads to counteracting effects that need to be balanced. For example, in order to be able to find the solution, a PQC needs to have many parameters. If there are too few parameters, there will be physically relevant states which cannot be expressed with the given PQC. However, many parameters means many gates and thus large noise. Being able to express all physically relevant states is therefore one measure for being a ``good'' PQC. Taking this point of view, an ``optimal'' PQC would not only be \emph{maximally expressive} (the circuit can generate all relevant states) but also \emph{minimal} (there are no redundant parameters, i.e., removing any parameter would reduce the set of states the PQC can generate). Analyzing PQCs from this point of view is the main objective of the dimensional expressivity analysis (DEA)~\cite{Funcke2021a,Funcke2021b}.

In these proceedings, we will review the theoretical background of the DEA in \autoref{sec:theory} and the hardware implementation in \autoref{sec:hardware}. In \autoref{sec:physical} we will review physical state spaces, which allows us to find a termination condition for PQC construction schemes that can have arbitrarily many parameters, and which allows us to automate custom circuit designs. In particular, the here presented automated circuit design explicitly extend the results of~\cite{Funcke2021a}. Finally, we will review best-approximation error estimates in \autoref{sec:best-approx} and provide some concluding remarks in \autoref{sec:conclusion}.

\section{Theoretical Background}\label{sec:theory}

In this section, we will discuss the theoretical background of the dimensional expressivity analysis. At the most fundamental level, the DEA aims to identify redundant parameters in a given PQC. To this end, we consider a \emph{parametric quantum circuit} to be a map $C$ from some parameter space $\Pp$ into the state space $\Sp$. The state space $\Sp$ may be the entire state space of the quantum device or just contain the physically relevant states. As such, the PQC includes the quantum device initialization, and for each set of parameters $\theta\in\Pp$, $C(\theta)$ is a state of the quantum device. For example,
  \begin{align*}
    \Qcircuit @C=1em @R=.7em {
      \lstick{\ket{0}} & \gate{R_Y(\theta_1)} & \gate{R_Z(\theta_3)} & \ctrl{1} & \gate{R_Y(\theta_5)} & \gate{R_Z(\theta_7)} & \qw \\
      \lstick{\ket{0}} & \gate{R_Y(\theta_2)} & \gate{R_Z(\theta_4)} & \targ    & \gate{R_Y(\theta_6)} & \gate{R_Z(\theta_8)} & \qw
    }
  \end{align*}
  is a PQC on a two-qubit device with initial state $\ket0\otimes\ket0$. The parametric gates in this circuit are rotation gates $R_G(\theta)=\exp\l(-\frac{i}{2}\theta G\r)$, where $G$ is a Pauli matrix $Y$ or $Z$ depending on the gate. For the purpose of these proceedings, we will restrict our considerations to such rotation gates $R_G$ with $G=X,Y,Z$. We refer to~\cite{Funcke2021a,Funcke2021b} for treatments of more general gates, such as $G=\CNOT(c,t)$ or $G=\sum_{q=0}^{Q-1}X_qX_{q+1}$ where $X_q$ is a Pauli $X$ gate acting on qubit $q$. In complete generality, any parametric dependence on the unitary operation performed on the quantum device can be considered.

\subsection{Identification of redundant parameters}\label{subsec:identify-params}
A parameter $\theta_k$ is considered to be \emph{redundant} if a small change of $\theta_k$ keeping all other $\theta_j$ fixed produces a state that can also be obtained by keeping $\theta_k$ fixed and adjusting all other $\theta_j$ accordingly. Geometrically speaking, this means that the partial derivative $\d_kC(\theta)$ of $C$ with respect to $\theta_k$ is a linear combination of the remaining partial derivatives $\d_jC(\theta)$. This can be checked inductively by considering the real partial Jacobians $J_k$ of~$C$,
\begin{align}
  J_k(\theta)=
  \begin{pmatrix}
    \Re\d_1C(\theta)&\cdots&\Re\d_kC(\theta)\\
    \Im\d_1C(\theta)&\cdots&\Im\d_kC(\theta)\\
  \end{pmatrix}.
\end{align}
As long as $\theta_1$ is non-trivial, $J_1$ will always have rank $1$. Adding the column for $\d_2C$, we can check whether $J_2$ has rank $2$. If that is the case, then we move on to the next parameter. If adding a column for $\d_kC$ does not increase the rank of $J_k$, then $\theta_k$ is redundant and can be set constant. This effectively removes the parameter and thus the corresponding column in $J_k$. Eventually, either all parameters have been checked and classified as independent/redundant or the number of independent parameters found equals the dimension of $\Sp$. In the latter case, we can stop the analysis as all further parameters must necessarily be dependent.

It should be noted that the choice of the \emph{real} partial Jacobian is necessitated by the fact that our parameter space $\Pp$ is commonly a real manifold. Thus, the linear dependence check has to be performed with respect to linear combinations in $\rn$ as opposed to $\cn$.

To check the rank of $J_k$ efficiently, we usually consider the matrix $S_k=J_k^*J_k$ instead. Since, by induction, we can assume that $J_{k-1}$ has full rank, $J_k$ has full rank if and only if all eigenvalues of $S_k$ are strictly positive. Thus, computing the smallest eigenvalue of $S_k$ (or otherwise checking $S_k$ for invertibility) only has a computational complexity in $\Op(k^3)$ and does not require working with the $J_k$ directly which are of dimension $2^{Q+1}\times k$ where $Q$ is the number of qubits. Alternatively, if the number of parameters $N$ for a given PQC is fixed (i.e., the PQC is not given by a construction that can use arbitrarily many parameters), computing the reduced row echelon form of $S_N$ can also identify all independent parameters.

\subsection{Two Bloch sphere examples}
To illustrate the analysis, let us consider two quantum circuits on a single qubit. The first circuit will have two independent parameters. The second circuit will have one dependent parameter.

\subsubsection{A minimal circuit}
The first circuit we want to consider is $C(\theta)=R_Z(\theta_2)R_X(\theta_1)\ket0$. Using $\ket0$ and $\ket1$ as a basis of the single-qubit Hilbert space, we can write
\begin{align}
  C(\theta)=&R_Z(\theta_2)R_X(\theta_1)\ket0
  =
  \begin{pmatrix}
    \cos\frac{\theta_1}{2}\cos\frac{\theta_2}{2}-i\cos\frac{\theta_1}{2}\sin\frac{\theta_2}{2}\\
    -i\sin\frac{\theta_1}{2}\cos\frac{\theta_2}{2}+\sin\frac{\theta_1}{2}\sin\frac{\theta_2}{2}
  \end{pmatrix}
  \begin{pmatrix}
    \ket0\\\ket 1
  \end{pmatrix},
\end{align}
which yields
\begin{align}
  J_1=\frac12
  \begin{pmatrix}
    -\sin\frac{\theta_1}{2}\cos\frac{\theta_2}{2}\\
    \cos\frac{\theta_1}{2}\sin\frac{\theta_2}{2}\\
    \sin\frac{\theta_1}{2}\sin\frac{\theta_2}{2}\\
    -\cos\frac{\theta_1}{2}\cos\frac{\theta_2}{2}
  \end{pmatrix}
  \qquad\text{and}\qquad
  J_2=\frac12
  \begin{pmatrix}
    -\sin\frac{\theta_1}{2}\cos\frac{\theta_2}{2}&-\cos\frac{\theta_1}{2}\sin\frac{\theta_2}{2}\\
    \cos\frac{\theta_1}{2}\sin\frac{\theta_2}{2}&\sin\frac{\theta_1}{2}\cos\frac{\theta_2}{2}\\
    \sin\frac{\theta_1}{2}\sin\frac{\theta_2}{2}&-\cos\frac{\theta_1}{2}\cos\frac{\theta_2}{2}\\
    -\cos\frac{\theta_1}{2}\cos\frac{\theta_2}{2}&\sin\frac{\theta_1}{2}\sin\frac{\theta_2}{2}
  \end{pmatrix}
\end{align}
and hence
\begin{align}
  S_1=J_1^*J_1 = \frac14\qquad\text{and}\qquad S_2=J_2^*J_2 = 
  \begin{pmatrix}
    \frac14&0\\0&\frac14
  \end{pmatrix}.
\end{align}
Both $S_1$ and $S_2$ are invertible. This implies that both parameters are independent. Since neither parameter can be removed without reducing the expressivity of the circuit, this circuit is minimal. 

\subsubsection{A reducible circuit}
If instead we consider the circuit $C(\theta)=R_X(\theta_2)R_X(\theta_1)\ket0$, then the same analysis yields 
\begin{align}
  S_1=J_1^*J_1 = \frac14\qquad\text{and}\qquad S_2=J_2^*J_2 = \frac14
  \begin{pmatrix}
    1&1\\1&1
  \end{pmatrix}.
\end{align}
Here, $S_1$ is invertible, i.e., the first $X$-rotation gate is independent. However, $S_2$ is not invertible and we have correctly identified that the second $X$-rotation has no contribution to the circuit that cannot already be achieved with the first $X$-rotation. Thus, $\theta_2$ is redundant and can be removed, that is, it can be set to a constant value. Depending on the circumstances, this constant value could be chosen to be $0$, thus removing the gate from the circuit, or it may be advantageous to keep $\theta_2$ at a non-trivial value. The latter would be more common in experimental setups where the dependent parameter may be a pulse duration and thus only certain values are experimentally viable, or this choice may be employed in on-the-fly circuit construction with seamless switching~\cite{Funcke2021a}.

\subsection{Removal of unwanted symmetries}\label{subsec:symmetries}
An important extension to the basic parameter identification is the ability to remove unwanted symmetries from a PQC. This may occur if a PQC should be constructed to be invariant under some symmetry that has no impact on the cost function, e.g., gauge invariance or more simply a global phase symmetry. As the symmetry has no impact on the cost function, removing any parameters that only contribute this symmetry would further improve the efficiency of many classical feedback loop optimizers used in the VQS. Moreover, this would likely reduce the amount of device noise because fewer gates need to be employed in the PQC.

To remove the unwanted symmetry, we need to formally extend the PQC $C(\theta)$ to a larger circuit $\tilde C(\phi,\theta)$ in such a way that (a) keeping $\theta$ fixed and varying $\phi$ only changes $\tilde C(\phi,\theta)$ by the action of the unwanted symmetry and (b) there exists a $\phi_0$ such that $\tilde C(\phi_0,\theta)=C(\theta)$.

For example, let us consider the circuit $C(\theta)=R_Y(\theta_3)R_Z(\theta_2)R_X(\theta_1)\ket0$. This circuit is minimal and can generate every single-qubit state. Thus, for every $\theta\in(\rn/2\pi\zn)^3$ and $\alpha\in U(1)$, there exists a $\theta'$ such that $C(\theta')=\alpha C(\theta)$. In this sense, $C$ has a global phase symmetry. To remove this symmetry, we extend $C$ to the circuit $\tilde C(\phi,\theta)=R_Y(\theta_3)R_Z(\theta_2)R_X(\theta_1)R_Z(\phi)\ket0$. Thus, keeping $\theta$ fixed and changing $\phi\to\phi'$ only produces a phase change $e^{-i\frac{\phi'-\phi}{2}}$ (property (a)) and $\tilde C(0,\theta)=C(\theta)$ (property (b)).

To finally remove the unwanted symmetry, we perform the DEA checking the parameters $\phi$ before $\theta$. This ensures that the unwanted symmetry is generated using $\phi$, and any parameter $\theta_k$ that only contributes the unwanted symmetry will now be dependent on $\phi$ and all $\theta_j$ with $j<k$. Setting $\phi$ to $\phi_0$ in the reduced circuit successfully removes the unwanted symmetry from the PQC.

In the case of $\tilde C(\phi,\theta)=R_Y(\theta_3)R_Z(\theta_2)R_X(\theta_1)R_Z(\phi)\ket0$, DEA shows that $\theta_3$ is dependent, i.e., we can reduce the circuit to $\tilde C_r(\phi,\theta)=R_Z(\theta_2)R_X(\theta_1)R_Z(\phi)\ket0$ which is still maximally expressive on a single qubit. Setting $\phi=0$ thus yields the reduced circuit $C_r(\theta)=R_Z(\theta_2)R_X(\theta_1)\ket0$ which can generate arbitrary single-qubit states up to a global phase. 

\section{Hardware Implementation}\label{sec:hardware}

If the DEA is to be used for quantum circuits on many qubits and even in on-the-fly circuit construction/optimization, then an efficient automation of the process is paramount. In \autoref{subsec:identify-params} we have noted that the $k\times k$ matrices $S_k=J_k^*J_k$ need to be checked for invertibility. For PQCs with $N$ parameters, this invertibility check can be performed for all $k$ with $\Op(N^2)$ memory requirements and $\Op(N^4)$ CPU calls. However, a classical computation of the $S_k$ requires $\Op(N\ 2^{Q+1})$ computational resources where $Q$ is the number of qubits. This is prohibitively expensive in the scaling limit. Thus, the computation of $S_k$ should make use of the quantum device.

Considering $S_k=J_k^*J_k$, we note that the $(m,n)$-element of $S_k$ is given by $\Re\langle\d_mC(\theta),\d_nC(\theta)\rangle$. Furthermore, if each parametric gate in $C$ is a rotation gate $R_{G_m}$, then $\d_mC(\theta)=-\frac{i}{2}\gamma_m\ket\init$, where $\gamma_m\ket{\init}$ is the circuit $C(\theta)$ with an additional gate $G_m$ inserted after $R_{G_m}(\theta_m)$, and $\ket\init$ is the initial state of the quantum device. Thus, the $(m,n)$-element of $S_k$ is given by $\frac14\Re\bra\init\gamma_m^*\gamma_n\ket\init$ and the computational objective becomes measuring $\Re\bra\init\gamma_m^*\gamma_n\ket\init$ on the quantum device. 

To measure $\Re\bra\init\gamma_m^*\gamma_n\ket\init$ on the quantum device, we can construct the state~\cite{Zhao-Zhao-Rebentrost-Fitzsimmons2019}
\begin{align}
  \ket{\psi_{m,n}}=\frac{\ket0\otimes\l(\gamma_m\ket{\init}+\gamma_n\ket{\init}\r)+\ket1\otimes\l(\gamma_m\ket{\init}-\gamma_n\ket{\init}\r)}{2}
\end{align}
since measuring the ancilla of $\ket{\psi_{m,n}}$ yields
\begin{align}
  \prob(\anc=\ket0)=\frac{1+\Re\bra{\init}\gamma_m^*\gamma_n\ket{\init}}{2}.
\end{align}
This can be achieved with the circuit 
\begin{align}\label{eq:circ-general-Re}
  \Qcircuit @C=1em @R=.7em {
    \lstick{\ket{\init}}                        & \qw         & \gate{\gamma_n}  & \qw         & \gate{\gamma_m} &\qw        & \qw         & \qw & \qw\\
    \lstick{\text{ancilla: }\ket{0}} & \gate{H}   &  \ctrl{-1}                & \gate{X}    & \ctrl{-1}                 & \gate{X} &\gate{H} & \meter &\cw\\
  }
\end{align}
using just one ancilla qubit. In fact, since $\gamma_m$ and $\gamma_n$ only differ by the application of $G_m$ after $R_{G_m}$ in $\gamma_m$ and $G_n$ after $R_{G_n}$ in $\gamma_n$, it suffices to apply the circuit $C(\theta)$ once and only use controlled~$G_m$ and controlled $G_n$ gates at the appropriate time. For example, if we consider $\Re\bra0 \gamma_2^*\gamma_1\ket0$ for $C(\theta)=R_Z(\theta_2)R_X(\theta_1)\ket0$, then the circuit~\eqref{eq:circ-general-Re} becomes
\begin{align}
  \Qcircuit @C=1em @R=.7em {
    \lstick{\ket{0}} & \gate{R_X(\theta_1)} & \gate{X} & \gate{R_Z(\theta_2)} & \gate{Z} &\qw        & \qw       & \qw & \qw\\
    \lstick{\ket{0}} & \gate{H}                    &  \ctrl{-1} & \gate{X}                    & \ctrl{-1} & \gate{X} &\gate{H} & \meter &\cw\\
  }
\end{align}
which requires only $\CNOT$ and $CZ$ gates and not the $CR_X$ and $CR_Z$ gates formally required by the circuit~\eqref{eq:circ-general-Re}.

\subsection{Experimental results}\label{subsec:experiment}

\begin{figure}[ht!]
  \includegraphics[width=\textwidth]{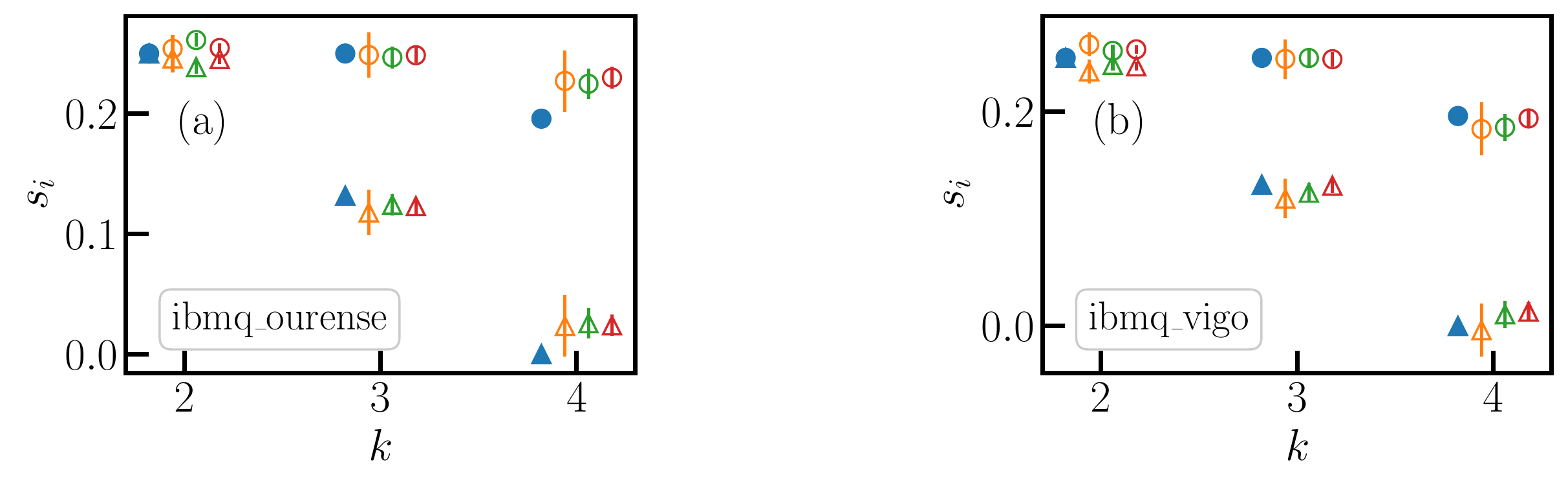}
  \caption{Single-qubit experiments on (a) ibmq\_ourense and (b) ibmq\_vigo quantum hardware. We show the smallest (triangles) and second smallest (dots) eigenvalues of the matrix $S_k$. The filled symbols are the exact solutions. The open symbols represent different statistics: orange, green and red markers stand for 1000, 4000 and 8000 shots, respectively. The error bars for the experimental data represent the uncertainty from the finite number of shots. This figure is discussed in more detail in~\cite{Funcke:2021aps}.
  }\label{fig:DEA-1-Q}
\end{figure}

In \autoref{fig:DEA-1-Q} we show the results of the hybrid quantum-classical DEA implementation for the single-qubit circuit $C(\theta)=R_Y(\theta_4)R_Z(\theta_3)R_X(\theta_2)R_Z(\theta_1)\ket0$ at a randomly selected value of $\theta$. We already know from \autoref{subsec:symmetries} that the parameters $\theta_1$, $\theta_2$, and $\theta_3$ are independent and $\theta_4$ is redundant. Thus, we expect to see that the smallest eigenvalue of $S_4$ is compatible with zero and all other eigenvalues are strictly positive. \autoref{fig:DEA-1-Q} shows the two smallest eigenvalues for $S_2$, $S_3$, and $S_4$ as computed using ibmq\_ourense and ibmq\_vigo as co-processors. In particular, \autoref{fig:DEA-1-Q} demonstrates that we can reliably identify $\theta_4$ as redundant and the other parameters as independent.

The error bars in \autoref{fig:DEA-1-Q} have been obtained using experimental reconstruction. Since the eigenvalues are computed from probabilities of finding the ancilla in $\ket0$, we do not have direct access to the error statistics of the matrices $S_k$. Instead, we used the observed $\prob(\anc=\ket0)$ and simulated $1000$ experimental outcomes with this underlying probability distribution. As such, we can estimate the standard deviation of the eigenvalues. This is shown as error bars in \autoref{fig:DEA-1-Q}.

\section{Physical state spaces}\label{sec:physical}
Since the quantum device state space grows exponentially with the number of qubits, having maximally expressive PQCs on the entire quantum device state space requires exponentially many parameters as well. This quickly leads to situations in which maximally expressive circuits are not computationally viable. However, in many physically relevant models, it is not necessary to have maximally expressive PQCs on the entire quantum device state space, since physical symmetries restrict the physically relevant state space to a dimension that grows only polynomially in the number of qubits. Thus, if we can construct a PQC that satisfies many of the physically relevant symmetries, then we only need polynomially many parameters and might be able to further reduce the number of parameters by removing any unwanted symmetries (cf., \autoref{subsec:symmetries}).

For example, if we consider translational symmetry, then any physical state has to be invariant under the translation operator $\tau_Q$ which maps the computational basis state $\ket{b_{Q-1}\ldots b_0}$ to $\ket{b_{Q-2}\ldots b_0b_{Q-1}}$, i.e., $\tau_3(\ket{101})=\ket{011}$. In general, we may consider a symmetry operator $\tau$ on $Q$ qubits. Then, the physical sectors are intersections of the eigenspaces of $\tau$ with the quantum device state space, i.e., the unit spheres in the eigenspaces of $\tau$.

To analyze such physical sectors~\cite{Funcke2021a}, we consider the action of $\tau$ on the computational basis and construct equivalence classes of states under the action of $\tau$. Hence, $\ket\psi\sim_\tau\ket\phi$ if $\ket\psi=\tau^j\ket\phi$ for some $j$. In case of translational symmetry,  $\ket{011}\sim_{\tau_3}\ket{101}$ but $\ket{011}\not\sim_{\tau_3}\ket{001}$. As such, we obtain equivalence classes $[\ket b]$ and the order of an equivalence class is the number of elements in the equivalence class, i.e., $[\ket{011}]=\{\ket{011},\ket{110},\ket{101}\}$ has order $3$. Similarly, the eigenvalues of $\tau$ have an order. For the translational invariance case, the eigenvalues are roots of unity $\omega$ and their order is the smallest $j\ge1$ with $\omega^j=1$. This implies that an equivalence class $[\ket b]$ can be mapped into the physical sector of an eigenvalue $\omega$ if the order of $\omega$ divides the order of $[\ket b]$. Explicitly, for translational invariance, the basis vectors of this mapping can be chosen to be
\begin{align}
  e(\omega,[\ket b]) = \frac{1}{\sqrt{\mathrm{order}([\ket b])}}\sum_{j=0}^{\mathrm{order}([\ket b])-1}\omega^j\tau_Q^j\ket b.
  \label{eq:mapping}
\end{align}
This allows us to compute the dimension of the physical sector corresponding to the eigenvalue $\omega$ with order $d$ on $Q$ qubits as~\cite{Funcke2021a}
\begin{align}\label{eq:dim_phys_sec}
  \dim_\rn(\text{Physical Sector of }\omega)=-1+2\sum_{d|k|Q}\frac{\#(k)}{k}
\end{align}
where $\#(1)=2$ and $\#(k)=2^k-\sum_{k'|k, k'<k}\#(k')$. 

\autoref{eq:dim_phys_sec} is also the number of independent parameters that a maximally expressive PQC mapping into this sector has. Hence, once this number of independent parameters is reached, all other parameters that a PQC may have must necessarily be redundant.

\subsection{Automatic custom circuit construction}\label{subsec:automatic}
In this section, we will extend the results of~\cite{Funcke2021a} to explicitly address the question of automating the custom circuit construction described in~\cite{Funcke2021a}. More precisely, the equivalence class construction can be exploited for automatic custom circuit design. We will first discuss how to construct the gates from the equivalence relations. As a second step, we will order the constructed gates based on how complicated they are to obtain an inductive automation process.

As an example, let us consider the case of the $\omega=1$ sector of translational invariance on $Q$~qubits. Based on Eq.~\eqref{eq:mapping}, we can write the states in this sector of the form $e_{[\ket B]}=e(1,[\ket B])=\{\mathrm{order}([\ket B])\}^{-1/2}\sum_{\ket b\in[\ket B]}\ket b$. If we consider the point $\ket0$, then we need to construct parametric gates whose derivatives at $\ket0$ point into the directions $e_{[\ket B]}$ both with a real and an imaginary coefficient. 

Starting with $\ket B\ne\ket0$, we first aim to construct a gate that generates a component in direction $e_{[\ket B]}$ with an imaginary coefficient. For this, we can consider the gate
\begin{align}
  R_{X^B}(\theta)=\exp\l(-i\frac{\theta}{2}\sum_{j=0}^{\mathrm{order}([\ket B])-1}X^{\tau_Q^j\ket B}\r),
  \label{eq:imagdir}
\end{align}
where $X^b=X_{Q-1}^{b_{Q-1}}\otimes\cdots\otimes X_0^{b_0}$. Then, $R_{X^B}'(0)\ket0=-\frac{i}{2}\,\l(\mathrm{order}([\ket B])\r)^{1/2}\,e_{[\ket B]}$, i.e., this gate generates a component in direction $ie_{[\ket B]}$ at $\ket0$.

As a next step, we aim to construct a gate that generates a component in direction $e_{[\ket B]}$ with a real coefficient. Since $B=B_{Q-1}\ldots B_0\ne0$, at least one of the $B_j$ is not $0$, and  we can choose one qubit $j$ and replace $X_j^{B_j}$ with $Y_j^{B_j}$. For example, for $B=101$, $B_0\ne0$ and $X^B=X_2\otimes X_0$ is changed to $X_2\otimes Y_0$. For the translational symmetry in \autoref{eq:imagdir}, $X_2\otimes Y_0$ is shifted in the qubit index through the action of $\tau_Q^j$, which creates operators $X_{j+2}\otimes Y_j$. Summing all these operators to replace the sum in \autoref{eq:imagdir} results in a gate $R_{(X|Y)^B}(\theta)$, e.g.,
\begin{align}
  R_{(X|Y)^{101}}(\theta)=\exp\l(-i\frac{\theta}{2}\l(X_2\otimes Y_0+X_0\otimes Y_1+X_1\otimes Y_2\r)\r). 
\end{align}
By replacing exactly one $X$ with $Y$, $R_{(X|Y)^B}$ generates a component in direction $e_{[\ket B]}$ at $\ket0$.

All these gates $R_{X^B}$ and  $R_{(X|Y)^B}$ for $B\ne0$ have independent parameters, as can be verified using DEA. The number of parameters is exactly one fewer than the dimension of the physical state space, and the missing gate at $\ket0$ should generate a component in the $i\ket0$ direction. This can be achieved with a translationally invariant $Z$-rotation gate
\begin{align}
  R_{Z,Q}(\theta)=\exp\l(-i\frac{\theta}{2}\sum_{q=0}^{Q-1} Z_q\r). 
\end{align}

Considering the set of gates $R_{Z,Q}$, $R_{X^B}$ and  $R_{(X|Y)^B}$, we can ``order'' them with respect to the number of $X$ or $Y$ gates acting simultaneously in each summand of the exponent in \autoref{eq:imagdir}. As such, $R_{Z,Q}$ is an order-$0$ gate and a gate of order $k$ corresponds to $R_{X^B}$ or $R_{(X|Y)^B}$, where $B$ has exactly $k$ values $B_j=1$. This allows for the process to be automated by inductively generating more complicated gates.  Using translational invariance and the fact that we are always summing over all elements of an equivalence class $[\ket B]$, we can choose a ``canonical'' representative $B$ of $[\ket B]$ by assuming $B_0=1$ and $B_{Q-1}=0$. This reduces the automation procedure to a combinatorial problem with complexity that is comparable to the dimensional scaling of the physical state space dimension. In other words, if the physical state space grows polynomially in the number of qubits, then this approach generates minimal and maximally expressive custom circuits with polynomially many parameters. 

\section{Best-approximation error}\label{sec:best-approx}
In some cases, having maximally expressive circuits may still be computationally too costly, even with targeted custom circuits that only generate physically relevant states. Thus, it is necessary to settle for non-maximally expressive PQCs. 
This means that some physically relevant states cannot be reached by the PQC $C$ and thus cannot be expressed as a state $C(\theta)$. In this case, we wish to compute the worst-case distance between a physically relevant $\ket\psi$ and the closest state~$C(\theta)$. This worst-case distance is the (worst-case) best-approximation error $\alpha_C$ of the circuit $C$.

In order to approximate $\alpha_C$, we generate a discrete sample set $\Dp$ of $N$ points $C(\theta_1),\ldots,C(\theta_N)$. If every point $C(\theta)$ is $\eps$-close to at least one point in $\Dp$, then we can compute the worst-case best-approximation error $\alpha_C^\Dp$ with respect to points in $\Dp$, rather than the entire image of $C$, and obtain $\alpha_C^\Dp-\eps\le\alpha_C\le\alpha_C^\Dp$~\cite{Funcke2021b}. Thus, if $\Dp$ becomes dense for $N\to\infty$, we obtain $\alpha_C^\Dp\to\alpha_C$, and for each $\epsilon$, we find a lower and upper bound on $\alpha_C$.
Since $\Dp$ is a finite discrete set, we can compute $\alpha_C^\Dp$ using Voronoi diagrams~\cite{Voronoi1908a,Voronoi1908b} in the physical state space. In particular, choosing $\Dp$ using a scrambled Sobol' sequence in parameter space shows a convergence of $\alpha_C^\Dp\to\alpha_C$ that is comparable to the theoretically optimal rate of convergence~\cite{Funcke2021b}.
However, since the Voronoi diagram computation is classical, we need an efficient mapping of the quantum states $C(\theta_j)$ in $\Dp$ into classical memory. We proposed~\cite{Funcke2021b} to do this by constructing a basis transformation from the linear space spanned by $\Dp$ to $\rn^{\dim_{\rn}\Sp+1}$, where $\Sp$ is the relevant state space. To this end, we need to compute $\Re\langle C(\theta_j),C(\theta_k\rangle$, which we can perform efficiently using the circuit \eqref{eq:circ-general-Re}. Thus, if the physically relevant state space has polynomial scaling in the number of qubits, then this hybrid quantum-classical algorithm of computing $\alpha_C^\Dp$  is as efficient as the Voronoi diagram computation.

Since computing Voronoi diagrams for large $\Dp$ still requires significant computational resources, it may be useful to have a more computationally efficient lower-bound estimate of~$\alpha_C$. If $C$ is known to be non-maximally expressive, then we can show $\alpha_C\gtrsim4\pi^{\frac{\dim\Mp}{2}+1}\Gamma\l(\frac{\dim\Mp}{2}\r)^{-1}\vol(\Mp)^{-1}$, where $\Mp$ is the image manifold of $C$~\cite{Funcke2021b}. Assuming $C$ is minimal (as can be ensured using DEA), $\dim\Mp$ is the number of parameters in $C$ and $\vol(\Mp)=\int_{\Pp}\sqrt{\det g(\theta)}d\vol_{\Pp}(\theta)$ can be computed using a suitable quadrature rule in the parameter space $\Pp$. Most importantly, $g$ is the same matrix as $S_N$ in \autoref{subsec:identify-params}, i.e., we already know an efficient hybrid quantum-classical algorithm to compute this lower-bound estimate on $\alpha_C$. Furthermore, we have observed that this lower-bound estimate is tight for spiral circuits on a single qubit~\cite{Funcke2021b}. 

\section{Conclusion}\label{sec:conclusion}
Given a parametric quantum circuit (PQC), DEA~\cite{Funcke2021a,Funcke2021b} allows us to identify redundant parameters in the PQC, i.e., parameters that do not increase the set of states the PQC can generate. This identification is based on an inductive procedure described in \autoref{subsec:identify-params}, which can be automated and efficiently implemented using the hybrid quantum-classical algorithm proposed in \autoref{sec:hardware}. The single-qubit experimental results of \autoref{subsec:experiment} and~\cite{Funcke2021a} demonstrate that we can reliably classify parameters. However, we have observed in~\cite{Funcke2021a} that current levels of hardware noise notably affect DEA on multiple qubits. It is therefore prudent to employ low-overhead error mitigation techniques if DEA is to be used for on-the-fly circuit construction/optimization.

For efficient use of the DEA, it is beneficial to start with a PQC design that takes certain physical symmetries into account (\autoref{sec:physical}) or is at least partially optimized ~\cite{Funcke2021b}. Then, the PQC can be further optimized by removing any unwanted symmetries (\autoref{subsec:symmetries}. A partially optimized starting point could be a minimal and maximally expressive circuit on the entire quantum device state space as proposed in~\cite{Funcke2021b}. 
Alternatively, physical symmetries can be used  for efficient termination conditions for parameter checks and 
automatic custom circuit construction (\autoref{sec:physical}).

Despite our optimization procedure, there may be cases in which it is be too computationally costly to have a maximally expressive PQC, e.g., due to hardware limitations. In these cases, a suitable candidate PQC can be chosen as a starting point and DEA can still provide minimal PQCs with the same expressivity the initial candidate PQC has. Furthermore, the DEA can be extended to include a priori estimates of the best-approximation error through the ideas outlined in \autoref{sec:best-approx}. In particular, lower bounds on the best-approximation error can be achieved with a computational overhead comparable to the DEA overhead for the given circuit itself. This can be extended to obtain asymptotically tight lower and upper bounds using a Voronoi diagram-based approach.

Of course, for current NISQ devices, the application range of DEA is still limited. To make DEA an integral part of PQC design, construction, and optimization, there are two main obstacles to overcome. On the one hand, hardware noise is still a limiting factor as we need to reliably identify the redundant and independent parameters. Thus, we need less noisy quantum hardware and low-overhead error mitigation techniques or even employ quantum error correction once the hardware is sufficiently advanced.
On the other hand, DEA only optimizes parametric gates and thus needs to be combined with non-parametric gate optimization techniques. The latter usually use different principles than DEA which makes it difficult to combine the methods. Thus, a unified approach of DEA and non-parametric optimization techniques would be a major step forward.

\section*{Acknowledgments}
  Research at Perimeter Institute is supported in part by the Government of Canada through the Department of Innovation, Science and Industry Canada and by the Province of Ontario through the Ministry of Colleges and Universities. L.F.\ is partially supported by the U.S.\ Department of Energy, Office of Science, National Quantum Information Science Research Centers, Co-design Center for Quantum Advantage (C$^2$QA) under contract number DE-SC0012704, by the DOE QuantiSED Consortium under subcontract number 675352, by the National Science Foundation under Cooperative Agreement PHY-2019786 (The NSF AI Institute for Artificial Intelligence and Fundamental Interactions, http://iaifi.org/), and by the U.S.\ Department of Energy, Office of Science, Office of Nuclear Physics under grant contract numbers DE-SC0011090 and DE-SC0021006. S.K.\ acknowledges financial support from the Cyprus Research and Innovation Foundation under project ``Future-proofing Scientific Applications for the Supercomputers of Tomorrow (FAST)'', contract no.\ COMPLEMENTARY/0916/0048. M.S.\ acknowledges the funding by the Helmholtz Einstein International Berlin Research School in Data Science (HEIBRiDS). P.S. acknowledges support from Agencia Estatal de Investigación (“Severo Ochoa” Center of Excellence CEX2019-000910-S, Plan National FIDEUA PID2019-106901GB-I00/10.13039 / 501100011033, FPI) ),, Fundació Privada Cellex, Fundació Mir-Puig, and from Generalitat de Catalunya (AGAUR Grant No. 2017 SGR 1341, CERCA program).

\bibliographystyle{JHEP}
\bibliography{Papers}


\end{document}